\documentclass[acmsmall,screen,nonacm]{acmart}
\setcopyright{none}
\usepackage{amsmath}
\usepackage{mathtools}                  
\usepackage{graphicx}
\usepackage{xcolor}
\usepackage{microtype}
\usepackage{pifont}                     
\usepackage{subcaption}                 
\usepackage{multirow}
\usepackage{booktabs}                   
\usepackage{colortbl}                   
\usepackage{arydshln}                   
\usepackage{tabularx}                   
\usepackage{tikz}
\usetikzlibrary{shapes.geometric, arrows.meta, positioning, decorations.pathreplacing, calc}
\usepackage{pgfplots}
\pgfplotsset{compat=1.18}
\usepackage{enumitem}
\usepackage[most]{tcolorbox}
\usepackage{listings}
\usepackage{algorithm}
\usepackage{algpseudocode}              
\usepackage{float}                      
\usepackage[capitalize,noabbrev]{cleveref}  
\crefname{figure}{Fig.}{Figs.}              
\Crefname{figure}{Fig.}{Figs.}

\definecolor{warmA}{HTML}{D6522A}      
\definecolor{warmAlight}{HTML}{F9D4C4} 
\definecolor{warmB}{HTML}{E07B50}      
\definecolor{warmBlight}{HTML}{FAE0D4} 
\definecolor{warmC}{HTML}{E8B84B}      
\definecolor{warmClight}{HTML}{FAF0CF} 
\definecolor{warmSuccess}{HTML}{6A9A50}      
\definecolor{warmSuccessLight}{HTML}{D2E8C3} 
\definecolor{warmFail}{HTML}{C84830}         
\definecolor{warmFailLight}{HTML}{F5C8BC}    
\definecolor{warmCream}{HTML}{FFF8EC}    
\definecolor{warmBorder}{HTML}{B08050}   

\definecolor{jsonstring}{HTML}{A31515}
\definecolor{jsonnumber}{HTML}{098658}
\definecolor{jsonbrace}{HTML}{0451A5}
\definecolor{jsonkey}{HTML}{0451A5}

\lstdefinelanguage{json}{
  basicstyle=\small\ttfamily,
  columns=fullflexible,
  keepspaces=true,
  showstringspaces=false,
  breaklines=true,
  literate=
    *{:}{{{\color{black}{:}}}}{1}
    {,}{{{\color{black}{,}}}}{1}
    {\{}{{{\color{jsonbrace}{\{}}}}{1}
    {\}}{{{\color{jsonbrace}{\}}}}}{1}
    {[}{{{\color{jsonbrace}{[}}}}{1}
    {]}{{{\color{jsonbrace}{]}}}}{1},
  string=[s]{"}{"},
  stringstyle=\color{jsonstring},
  morecomment=[l]{//},
  commentstyle=\color{gray}\itshape,
}

\newtcolorbox{rqsummarybox}[1][]{%
  enhanced,
  colback=warmCream, colframe=warmBorder,
  coltitle=warmBorder!25!black,
  fonttitle=\bfseries\small\sffamily,
  title={#1},
  rounded corners, arc=2mm,
  left=5pt, right=5pt, top=3pt, bottom=3pt,
  boxrule=0.6pt,
  shadow={1mm}{-1mm}{0mm}{warmBorder!15!white},
}

\definecolor{findingFrame}{HTML}{B8B8E6}
\definecolor{findingBack}{HTML}{F3F3FC}
\newtcolorbox{findingbox}[1]{
  colback=findingBack,
  colframe=findingFrame,
  colbacktitle=findingFrame,
  coltitle=black,
  fonttitle=\bfseries,
  title={#1},
  sharp corners,
  boxrule=0.8pt,
  leftrule=3pt,
  titlerule=0.4pt,
  arc=0mm,
  left=5pt,
  right=5pt,
  top=3pt,
  bottom=3pt,
  before skip=6pt,
  after skip=6pt
}

\newcommand{\hcell}[1]{%
  \ifnum#1>59\cellcolor{warmFail!50}\else
  \ifnum#1>29\cellcolor{warmFail!26}\else
  \ifnum#1>0\cellcolor{warmFail!9}\fi\fi\fi #1}

\definecolor{logGrey}{HTML}{D9D9D9}  
\newtcolorbox{systempromptbox}[1][]{%
  enhanced, breakable,
  colback=logGrey!18, colframe=logGrey, coltitle=black,
  fonttitle=\bfseries\small\sffamily,
  title={#1},
  rounded corners, arc=3mm,
  left=4pt, right=4pt, top=2pt, bottom=2pt,
  boxrule=0.6pt,
  fontupper=\small\ttfamily,
  before upper={\parindent=0pt\parskip=0.4em\obeylines},
}
\newtcolorbox{userpromptbox}[1][]{%
  enhanced, breakable,
  colback=green!3, colframe=green!40!black,
  fonttitle=\bfseries\small\sffamily,
  title={#1},
  rounded corners, arc=3mm,
  left=4pt, right=4pt, top=2pt, bottom=2pt,
  boxrule=0.6pt,
  fontupper=\small\ttfamily,
  before upper={\parindent=0pt\parskip=0.4em\obeylines},
}

\newcommand{\acAtOneThinkdepthaiQwen}{79.2\%}

\newcommand{\acAtOneClaudecodeQwen}{79.6\%}

\newcommand{\acAtOneTaskweaverQwen}{65.2\%}

\newcommand{\acAtOneMabcQwen}{42.6\%}

\newcommand{\acAtOneAiqQwen}{77.6\%}

\newcommand{\acAtOneThinkdepthaiSonnet}{90.0\%}

\newcommand{\acAtOneOpenrcaQwen}{46.6\%}

\newcommand{\faultGrandTotal}{500}

\begin{document}
\raggedbottom

\title{Beyond Fault Localization: A Trajectory-Level Study of LLM Agents for Microservice Root Cause Analysis}

\author{Qisheng Lu}
\affiliation{\institution{The Chinese University of Hong Kong, Shenzhen}\city{Shenzhen}\country{China}}
\email{qishenglu@link.cuhk.edu.cn}
\author{Aoyang Fang}
\authornote{Pinjia He and Aoyang Fang are co-corresponding authors.}
\affiliation{\institution{The Chinese University of Hong Kong, Shenzhen}\city{Shenzhen}\country{China}}
\email{aoyangfang@link.cuhk.edu.cn}
\author{Junjielong Xu}
\affiliation{\institution{The Chinese University of Hong Kong, Shenzhen}\city{Shenzhen}\country{China}}
\email{siyuexi@foxmail.com}
\author{Jin'ao Shang}
\affiliation{\institution{Xi'an Jiaotong University}\city{Xi'an}\country{China}}
\email{jinao\_s@stu.xjtu.edu.cn}
\author{Songhan Zhang}
\affiliation{\institution{The Chinese University of Hong Kong, Shenzhen}\city{Shenzhen}\country{China}}
\email{songhanzhang@link.cuhk.edu.cn}
\author{Yifan Yang}
\affiliation{\institution{The Chinese University of Hong Kong, Shenzhen}\city{Shenzhen}\country{China}}
\email{yifanyang6@link.cuhk.edu.cn}
\author{Xiaochuan Yan}
\affiliation{\institution{The Chinese University of Hong Kong, Shenzhen}\city{Shenzhen}\country{China}}
\email{xiaochuanyan@link.cuhk.edu.cn}
\author{Pinjia He}
\authornotemark[1]
\affiliation{\institution{The Chinese University of Hong Kong, Shenzhen}\city{Shenzhen}\country{China}}
\email{hepinjia@cuhk.edu.cn}

\begin{abstract}
Automated microservice root cause analysis (RCA) is usually evaluated by determining whether the final answer localizes the fault source.
However, this kind of evaluation is insufficient for on-call site reliability engineers (SREs), who need the evidence and reasoning behind the fault.
Evaluating the diagnostic process is challenging because agent trajectories interleave free-form natural-language reasoning with heterogeneous tool calls and telemetry results, making it difficult to assess agents' behavior.

To assess how well RCA agents ground their diagnoses in evidence, we propose a trajectory-analysis methodology that characterizes evidence navigation and utilization, SRE-oriented diagnostic intent, and recurring failure modes.
Our key insight is that fault-propagation paths provide a causal reference for assessing where agents investigate and how they use evidence.
Using this methodology, we analyze 5{,}000 trajectories from ten framework--backbone configurations on RCABench, covering two backbone models and nine agent frameworks.
We find that correct localization can hide deficiencies in causal reconstruction.
Successful investigations use retrieved evidence to guide subsequent queries, consider a broader range of diagnostic intents, and investigate the problem more deeply.
Failed investigations drift toward unrelated services or repeat shallow queries, with recurring failure patterns arising from omitted evidence, misinterpreted evidence, and unsupported reasoning.

To test whether these mined failure patterns can guide improvement, we instantiate a lightweight defense prototype targeting the three failure families.
Across two held-out benchmarks, two backbone models, and six diagnostic frameworks, the cataloged failure patterns recur; the prototype reduces their prevalence and raises $\mathrm{Acc}@1$ by 3.5--35.1 percentage points across all tested configurations, suggesting that the mined failures both generalize and are actionable.
Taken together, our findings motivate a process-level view of automated RCA, in which diagnostic trajectories reveal whether a diagnosis is not only correct, but also well-grounded and open to improvement.

\end{abstract}


\maketitle

\section{Introduction}\label{sec:introduction}

Microservice architectures underpin large-scale cloud applications~\cite{Guo2020GraphTrace}, where local faults can propagate across loosely coupled services and trigger cascading failures.
When such incidents occur, on-call site reliability engineers (SREs) must diagnose the fault source from distributed traces, logs, and metrics before remediation can begin~\cite{GoogleSRE2016Troubleshooting}.
Because remediation acts on a diagnosis, the SRE needs not only a root-cause label but also the evidence and reasoning that make it trustworthy.

\begin{figure}[t]
\centering
\includegraphics[width=\linewidth]{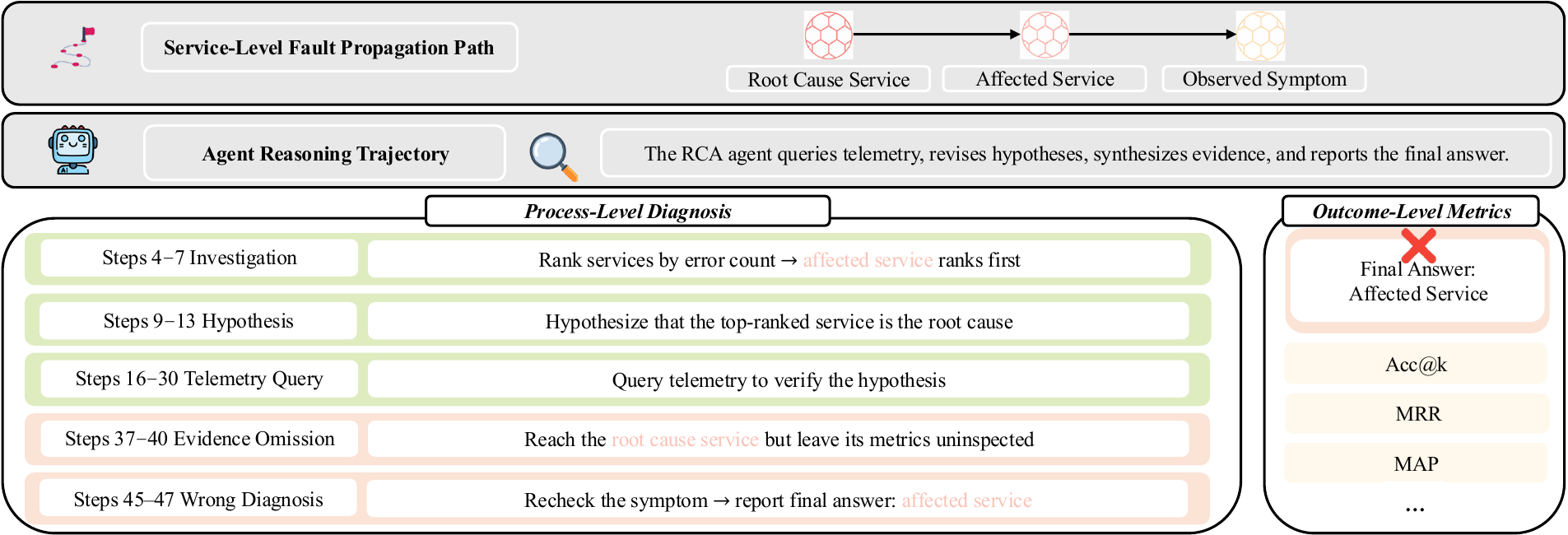}
\caption{Outcome-level metrics (right) record only the final answer and its correctness score. The diagnostic trajectory (left) exposes the process behind the same misdiagnosis: the agent navigates to the root cause service (Steps 37--40) but leaves its metrics uninspected, then rechecks the symptom and reports the affected service as its answer. The annotated fault propagation path (top) provides the process-level ground truth against which such failures can be identified.}
\label{fig:blackbox-vs-whitebox}
\end{figure}

Automated root cause analysis (RCA) has produced a large body of methods~\cite{irclsurvey, Zhang2024FailureDiagnosis, Fang2025GoalDrivenSurvey}.
Benchmark-driven studies usually turn RCA into a service-localization task: predict which service caused the incident.
Top-$k$ localization accuracy ($\mathrm{Acc}@k$), mean reciprocal rank, and mean average precision make this task comparable across methods, but they collapse the diagnostic process into a final label~\cite{surveyrcatsc, fang2026openrca}.
That collapse matters because the same correct label can be supported by strong or weak evidence, and a wrong label is only useful for continuous improvement if its failed reasoning can be inspected (\Cref{fig:blackbox-vs-whitebox}).
When such reasoning failures recur across incidents, they form reusable failure modes that can feed back into the diagnostic process.

Tool-using LLM systems make this missing diagnostic process newly visible.
General-purpose coding and deep-research agents now plan over many steps, invoke tools, and synthesize evidence~\cite{bouzenia2025understanding, du2025deepresearch}.
LLM-based RCA methods follow the same pattern by querying telemetry and revising hypotheses during diagnosis~\cite{chen2024automatic, coca, zhang2024mabc, xu2025openrca, luo2026graph, zhang2026towards, luo2025opsagent, tian2026gala, gao2026rcaflow, shi2025flowxpert}.
Because these agents act through observable steps, their runs leave trajectories: records of agent messages, telemetry queries, and tool-returned evidence.
Such trajectories can expose how an agent investigates an incident, not only what answer it returns.

However, raw trajectories cannot be evaluated directly as process-level evidence.
Prior RCA work studies causal graphs and fault propagation for localization~\cite{irclsurvey, Zhang2024FailureDiagnosis, Fang2025GoalDrivenSurvey}, and agent studies analyze tool traces~\cite{bouzenia2025understanding, du2025deepresearch}, but causal graphs capture statistically inferred service dependencies, not the incident-specific route a fault actually traveled, and agent-trace analyses lack a domain-level ground truth to evaluate against.
Evaluating diagnostic trajectories in RCA therefore requires a reference for what the diagnostic process should have recovered.
For our setting, that reference is a manually validated service-level fault propagation path: the route by which a fault moves from the faulty service to the observed symptom through service-dependency hops.
RCABench~\cite{fang2025rethinking}, the microservice benchmark used for our main study, provides final root-cause labels but not these process-level annotations.
Even with such paths, trajectories still need normalization because heterogeneous agents expose different trace formats and tool interfaces.
Without both process-level ground truth and normalized steps, existing metrics cannot say whether an agent navigated relevant evidence, what operational purpose each step served, or why wrong diagnoses recurred.

We address this gap with a trajectory-level analysis methodology built around service-level fault propagation paths.
These paths define the fault-impact surface: the services and edges through which a fault reaches the observed symptoms.
To compare agent behavior against this surface, we normalize each run into agent messages, telemetry queries, and tool-returned evidence.
The methodology applies outcome-level metrics to all nine frameworks and adds step-level trajectory analysis where the shared telemetry interface makes tool traces comparable.
The resulting analysis first checks whether a run finds the root service and reconstructs the propagation path.
The analysis then explains the route the agent took by mapping how the agent navigates the fault-impact surface, labeling SRE diagnostic intent and coding the failure modes behind wrong diagnoses.
Together, these steps make RCA trajectories comparable across frameworks and suitable for mining recurring diagnostic failures.

With this methodology in place, the study pursues two questions.
The first is \emph{characterization}: what separates a sound diagnostic process from a failing one?
We pursue it at three progressively deeper levels (RQ1--RQ3), moving from the outcome and cost of localization, to how an agent navigates the fault-impact surface and what diagnostic intent drives each query, to why wrong diagnoses occur.
We compare nine frameworks under a shared backbone model and isolate the backbone-model effect by rerunning one framework on a second model.
The trajectory-level analysis (RQ2--RQ3) focuses on the four frameworks whose tool calls are SQL queries against a shared telemetry store, because they expose comparable action traces and, as the stronger performers, yield the most informative diagnostic trajectories to analyze.
The second is \emph{intervention and validation} (RQ4--RQ5):
do the characterized failure modes recur beyond the original
study setting, and can they guide effective interventions?
We use these failure modes to design \textsc{DiagGuard},
a lightweight defense prototype that augments the diagnostic
agent with a grounding defense (the \emph{Grounder}) and
a verification defense (the \emph{Verifier}).
We examine failure-mode recurrence and changes in prevalence,
evaluate whether \textsc{DiagGuard} improves localization across
datasets, backbone models, and frameworks, and assess how
the two defenses contribute individually and jointly.

Our study yields three core findings. 1) Localization difficulty is uneven, accuracy falls as the causal chain deepens, and reconstructing how a fault propagates between services is harder than identifying which services it reaches, even when the fault source is correctly localized.
2) Success and failure hinge on the diagnostic process itself: effective runs stay on the fault-impact surface and act on the evidence they retrieve, and the breadth of their intent repertoire tracks how deep that search reaches; failing runs drift off the surface, leave decisive evidence unused, or circle in shallow loops.
3) Wrong diagnoses collapse into a small, evidence-handling failure taxonomy; an intervention distilled from these modes raises localization accuracy across datasets, backbone models, and frameworks and reduces the modes' prevalence, while the composition of failure modes remains more similar across backbones within the same framework than across frameworks.

This paper makes the following contributions:

\begin{itemize}[leftmargin=*,nosep]
  \item \textbf{Annotated propagation-path dataset.} We annotate and manually validate service-level fault propagation paths for a microservice RCA benchmark, supplying process-level ground truth beyond final root-cause labels.
  \item \textbf{Trajectory-level analysis methodology.} We develop a propagation-path-grounded methodology that normalizes heterogeneous RCA trajectories against annotated propagation paths, with adapted skills for mining diagnostic failure modes across diverse LLM-based agent frameworks.
  \item \textbf{Process-level empirical study.} We analyze 5{,}000 reasoning trajectories across ten framework--backbone configurations (nine frameworks under a shared backbone model, one rerun on a second model), linking diagnostic behavior and failure modes to final RCA outcomes.
  \item \textbf{Failure-mode-derived intervention.} We use the mined failure modes to design \textsc{DiagGuard} and validate it across datasets, backbone models, and frameworks, with failure-mode analysis on 1{,}368 additional trajectories confirming that the catalogued modes recur and the intervention generalizes beyond the original characterization setting.
\end{itemize}

We will release all artifacts, including the annotated propagation-path dataset, trajectory-analysis pipeline, failure-mode codebook, and \textsc{DiagGuard} prompts.

\section{Related Work}\label{sec:related_work}

We review traditional and LLM-based RCA (\Cref{sec:rw_rca,sec:rw_llm_rca}), the general-purpose agents we repurpose for RCA (\Cref{sec:rw_advanced_agents}), and trajectory-analysis techniques (\Cref{sec:rw_trajectory}). Across these lines, general-purpose deep-research and coding agents have not been systematically compared with RCA-specific designs on end-to-end RCA, and diagnostic processes are either not exposed or not evaluated against process-level ground truth.

\subsection{Traditional RCA in Microservice Systems}\label{sec:rw_rca}

Automated RCA research spans four families of methods: statistical correlation~\cite{Luo2019LatentError, epsilondiagnosis}, graph-based fault propagation~\cite{Wu2020MicroRCA, microrank}, causal inference~\cite{Azam2023CausIL, circa, torai, gao2025framework, zhang2025dynacausal}, and learning-based frameworks over multimodal observability data~\cite{Lee2023Eadro, Yu2023Nezha, Zheng2024MULAN}.
These methods return a root-cause ranking rather than a reasoning process; LLM-based and agent-based methods have recently emerged as a more process-oriented direction~\cite{Fang2025GoalDrivenSurvey}.

\subsection{LLM-Based RCA for Microservices}\label{sec:rw_llm_rca}

The application of large language models to microservice RCA has produced two distinct lines of work, following the \emph{workflow}-versus-\emph{agent} distinction~\cite{anthropic2024agents}.

\textit{Workflow approaches.}
These methods preprocess multimodal telemetry into a structured prompt for single-pass inference. They target classification and confidence estimation~\cite{zhang2024lm, chen2024automatic, han2024potential} or explanation and remediation~\cite{zhang2024automated, ahmed2023recommending, goel2024x, hrusto2024autonomous, sun2025multimodal}, or enrich the prompt with structured knowledge such as mined code or offline causal graphs~\cite{coca, metarca}.
A single snapshot of system state, however, cannot capture the progressive, hypothesis-driven nature of expert fault diagnosis.

\textit{Agentic approaches.}
These methods reason over multi-turn loops of diagnostic tool calls, telemetry queries, and hypothesis refinement, and differ in how they organize the diagnostic search.
\emph{Iterative diagnosis} agents drive a ReAct-style loop over telemetry, refining predictions through self-consistency or alternative prompting~\cite{xu2025openrca, rcagent, roy2024exploring}.
\emph{Graph-structured reasoning} constrains exploration with belief graphs, evidence graphs, or service-dependency graphs~\cite{luo2026graph, zhang2026towards, tian2026gala, amerrcl, zhang2026hypothesize}.
\emph{Multi-agent orchestration} coordinates multiple agents or encodes operational procedures via consensus mechanisms, behavior trees, or state machines~\cite{zhang2024mabc, gao2026rcaflow, pei2025flow, stratus, trioxpert, tamo}.
\emph{Knowledge and experience accumulation} reuses historical diagnostic knowledge through RAG, graph knowledge bases, or self-evolving skills~\cite{luo2025opsagent, shi2025flowxpert, sun2026opsmem, huang2026general}.
\emph{RCA-specific fine-tuning} trains the backbone on diagnostic data via SFT and RL~\cite{foundroot, thinkfl}; unlike the off-the-shelf agents our study examines, these require task-specific training.

These methods are nonetheless assessed almost exclusively at the outcome level, leaving the diagnostic process unexamined for lack of process-level ground truth.

\subsection{Advanced Agents}\label{sec:rw_advanced_agents}

General-purpose deep-research and coding agents now match or exceed domain-specific methods on knowledge-intensive tasks, raising the question of whether investigative architecture matters more than domain specialization.
The DeepResearch Bench~\cite{du2025deepresearch, li2026deepresearch} provides the first systematic evaluation of deep-research agents (DRAs); on the open-source side, \textsc{AIQ}~\cite{nvidiaaiq2025}, \textsc{TaskWeaver}~\cite{qiao2023taskweaver}, and \textsc{ThinkDepth.ai}~\cite{thinkdepthai2025} are representative frameworks, which we repurpose for RCA under the same harness (\Cref{sec:method_design}).
Commercial coding agents such as \textsc{ClaudeCode} (Anthropic's Claude Code) integrate with development environments to navigate codebases and iterate on solutions; we adapt one to RCA through the same harness, comparing it directly with RCA-specific agents.

\subsection{Agent Trajectory Analysis}\label{sec:rw_trajectory}

Inspecting agent trajectories has attracted growing interest: tools visualize and debug agent behavior~\cite{lu2024agentlens, epperson2025interactive}, and a parallel line mines historical trajectories to optimize agents~\cite{gupta2024metareflection, song2024trial, deng2024novice}.
For failure understanding, Bouzenia and Pradel~\cite{bouzenia2025understanding} and Zhang et al.~\cite{zhang2025agent} analyze trajectories of software-engineering and multi-agent systems.
Closest to ours, three concurrent efforts target RCA agents: Riddell et al.~\cite{forge2026} study post-detection RCA, where externally extracted anomalies are analyzed by agents over a static system graph, Kim et al.~\cite{kim2026whyrca} diagnose trajectories from a single \textsc{OpenRCA} architecture across multiple models, and STAR~\cite{star2026} repairs failing diagnostic stages post-hoc. Independently, OpenRCA~2.0~\cite{fang2026openrca} annotates step-level causal processes and reports that 76.0\% of agents identify a correct root-cause service yet only 61.5\% follow a verified propagation path, mirroring the Node\,F1 versus Edge\,F1 gap observed in our study. We instead evaluate the end-to-end RCA capability of multiple real agent frameworks and backbone models directly over telemetry against manually annotated propagation-path ground truth and turn the mined failure modes into a preventive intervention (\Cref{sec:treatment}).

\section{Study Setup}\label{sec:prelim}

\begin{figure*}[t]
\centering
\includegraphics[width=\textwidth]{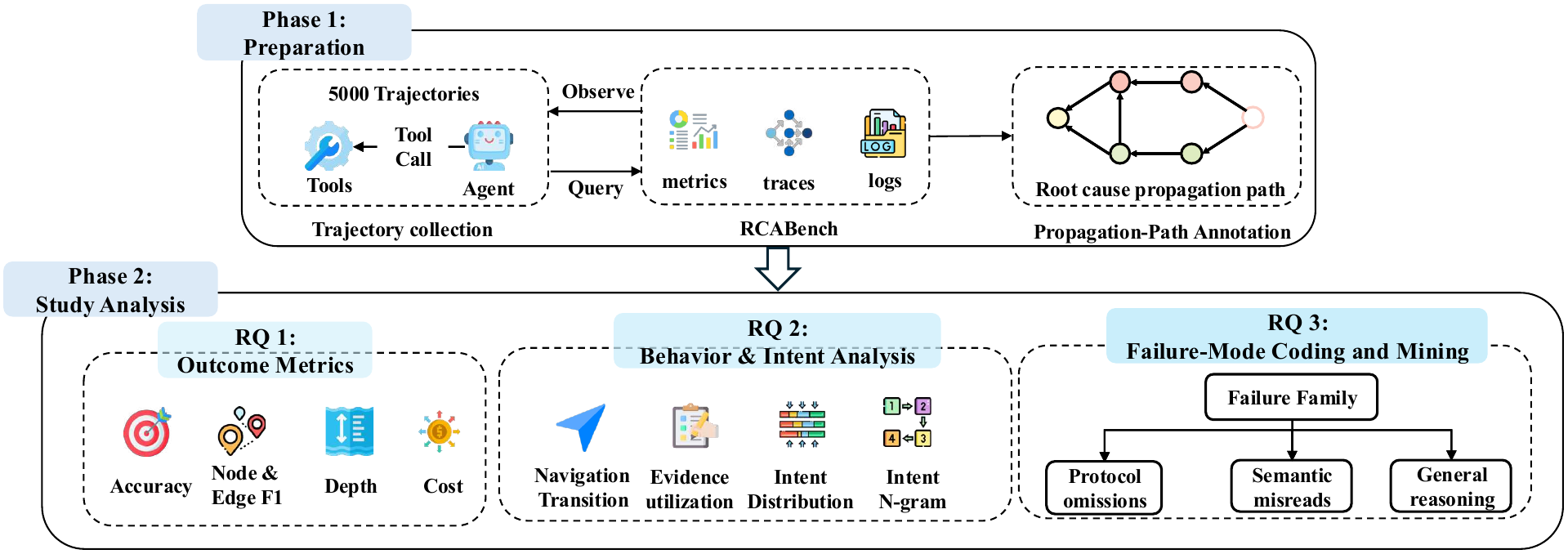}
\caption{\textbf{Study overview.} \emph{Phase~1} collects 5{,}000 agent reasoning
trajectories on RCABench: each agent queries telemetry through tool calls, and we annotate
every case's fault propagation path. \emph{Phase~2} characterizes them at
three deepening levels: RQ1 outcome and cost (\Cref{sec:rq1}), RQ2 behavior and
intent (\Cref{sec:rq2}), and RQ3 failure-mode families (\Cref{sec:rq3}); and
\textsc{DiagGuard} (RQ4--RQ5) appears in \Cref{fig:treatment}.}
\label{fig:overview}
\end{figure*}

This section establishes the study setup used throughout the paper: the evaluated agents, backbone models, and benchmarks. \Cref{fig:overview} gives the study design and the RQ1--RQ3 analysis pipeline; the validation (RQ4--RQ5) follows in \Cref{sec:treatment}.
\subsection{Evaluated Agent Frameworks}\label{sec:method_design}
To characterize the end-to-end RCA capability of LLM-based agent frameworks and analyze their diagnostic reasoning trajectories, we select nine frameworks spanning two categories (see \Cref{sec:rw_llm_rca} for design details).
The \emph{general-purpose} group includes four top-ranked investigative agents: \textsc{ThinkDepth.ai}~\cite{thinkdepthai2025}, \textsc{AIQ}~\cite{nvidiaaiq2025}, \textsc{TaskWeaver}~\cite{qiao2023taskweaver}, and \textsc{ClaudeCode}.
The \emph{RCA-specific} group includes five representative designs with publicly available code: \textsc{OpenRCA}~\cite{xu2025openrca}, \textsc{mABC}~\cite{zhang2024mabc}, \textsc{GoS}~\cite{luo2026graph}, \textsc{RCLAgent~V2}~\cite{zhang2026towards}, and \textsc{OpsAgent}~\cite{luo2025opsagent}, spanning iterative diagnosis, graph-structured reasoning, multi-agent orchestration, and experience accumulation.
All nine use off-the-shelf backbone models; fine-tuning approaches (\Cref{sec:rw_llm_rca}) are excluded because our study targets the reasoning process of general-purpose LLMs.

\subsection{Backbone Models}\label{sec:prelim_models}
Our study uses the following three backbone models (USD per 1M input/output tokens, as of 2026-02-15):
\begin{itemize}[leftmargin=*,nosep]
  \item \texttt{qwen3.5-plus-2026-02-15} (Qwen): \$0.115/\$0.688.
  \item \texttt{claude-sonnet-4.6} (Sonnet): \$3/\$15.
  \item \texttt{doubao-seed-2.0-pro-2026-02-15} (Doubao): \$0.44/\$2.22.
\end{itemize}

\subsection{Benchmarks}\label{sec:method_data}
We use three public benchmarks with different topologies and fault granularities.
\begin{itemize}[leftmargin=*,nosep]
  \item RCABench~\cite{fang2025rethinking}: a microservice RCA benchmark with
  hierarchical service-to-code-level fault labels; we evaluate on a
  \faultGrandTotal{}-case sample drawn by simple random sampling without
  replacement from its 1{,}430 cases.
  \item AIOps~2025~\cite{cai2026multi}: a microservices e-commerce RCA
  benchmark of 400 incidents over ten core services.
  \item RCA100~\cite{cai2026multi}: 103 chaos-drill incidents on the OpenTelemetry Demo; we use the 57-case propagation subset, i.e., the cases whose service-level propagation path, derived from the benchmark's annotated fault chain and the per-case topology, has at least one hop and is supported hop by hop by the released telemetry.
\end{itemize}

On RCABench, the empirical study (RQ1--RQ3) runs a horizontal comparison of all nine
frameworks under the shared Qwen and a vertical comparison of Qwen
against Sonnet under \textsc{ThinkDepth.ai} to isolate the backbone-model effect,
giving ten framework--backbone configurations (arms). The RQ2--RQ3 trajectory-level analysis focuses on the four general-purpose SQL-call frameworks; the five RCA-specific frameworks retain their native data access and enter the RQ1 baseline comparison without changing the trajectory-analysis cohort. In the intervention and validation phase (RQ4--RQ5), we examine
whether the mined failure modes recur beyond the characterization
setting and whether \textsc{DiagGuard}, the prototype derived from
these modes, improves localization.
On the 57-case propagation subset of RCA100, we evaluate
baseline and \textsc{DiagGuard}-augmented versions of six frameworks
(five RCA-specific plus \textsc{ThinkDepth.ai}) under both
Qwen and Doubao, and analyze their failure modes.
On the 400 AIOps~2025 incidents, we evaluate
\textsc{ThinkDepth.ai} with and without \textsc{DiagGuard}
under both backbone models.
We further ablate the two defenses using \textsc{ThinkDepth.ai}
with Doubao on AIOps~2025.

\section{Characterization Methodology}\label{sec:methodology}

This section details how we turn heterogeneous agent runs into path-grounded,
comparable measurements: annotating ground-truth propagation paths and collecting
trajectories (\Cref{sec:method_annotation}), defining outcome and difficulty
metrics (\Cref{sec:method_defs}), and analyzing trajectory behavior, intent, and
failure modes (\Cref{sec:method_behavior}, \Cref{sec:method_repair}).

\subsection{Propagation-Path Annotation and Trajectory Collection}\label{sec:method_annotation}

\paragraph{Propagation-path annotation.}
RCABench labels each case's root cause but not the service-level path to the SLO-alarming service, so we annotate this process-level ground truth ourselves. From
the injection metadata (targeted service and fault window), the pre-/post-injection
traces, metrics, and logs, and the static TrainTicket service-call
graph~\cite{trainticket}, we reconstruct a service-level fault propagation graph
$G^{\star}=(V^{\star},E^{\star})$, the union of one or more propagation paths from the root cause to the symptom. Its nodes are
the fault-affected services; each directed edge is a propagation step, realized
through a logical service call or physical resource sharing and grounded in the
abnormal spans, metric shifts, or log signatures within the fault window. This
per-case $G^{\star}$ is the ground truth for Node\,F1 and Edge\,F1
(\Cref{sec:method_defs}).

\paragraph{Trajectory collection.}
We collect $5{,}000$ reasoning trajectories. Each run yields thought-action-result
triples; a final \emph{compression prompt} then instructs the agent to reason over the collected evidence and report, in a uniform format, a predicted root-cause service and a self-reported causal propagation graph $G=(V,E)$, whose edges need not be query-verified. Both ground the outcome metrics (\Cref{sec:method_defs}).

\subsection{Metrics Definitions}\label{sec:method_defs}

\paragraph{Trajectory sequence.}
For each case, the agent's reasoning is an ordered sequence of \emph{rounds}
$\mathcal{T} = [\tau_1, \dots, \tau_n]$, each a triple
$\tau_i = (t_i, a_i, r_i)$ of thought, action, and result. The four general-purpose frameworks generate SQL
over a shared telemetry interface; the five RCA-specific ones
(\textsc{mABC}, \textsc{OpenRCA}, \textsc{GoS}, \textsc{OpsAgent},
\textsc{RCLAgent~V2}) retain their native data access.

\paragraph{Reasoning budget.}
We report each run's per-case reasoning budget as three quantities: \emph{rds}
(\emph{effective rounds}: rounds issuing at least one tool call), \emph{ktok}
(mean input+output tokens, in thousands), and \emph{\$} (those tokens priced at
the backbone model's list price; \Cref{sec:prelim_models}).

\paragraph{Root-cause localization.}
Each case has a ground-truth root-cause set $R^{\star}_i$ of one or two services.
$\mathrm{Acc}@1$ scores a case correct when the agent's predicted root cause
$\hat{r}_i$ lies in $R^{\star}_i$:
\begin{equation}\label{eq:acc_at_k}
  \mathrm{Acc}@1 = \frac{1}{N}\sum_{i=1}^{N}
    \mathbf{1}\!\left(\hat{r}_i \in R^{\star}_i\right),
\end{equation}
where $N$ is the number of incidents.
For configurations run multiple times (\Cref{sec:ablation}),
we also report $\mathrm{pass}@k$, the fraction of incidents with
$\hat{r}^{(j)}_i \in R^{\star}_i$ in at least one of $k$ runs:
\begin{equation}\label{eq:pass_at_k}
  \mathrm{pass}@k = \frac{1}{N}\sum_{i=1}^{N}
    \mathbf{1}\!\left(\exists\, j \in \{1,\dots,k\} : \hat{r}^{(j)}_i \in R^{\star}_i\right).
\end{equation}

\paragraph{Node and edge F1.}
Let $G^{\star}=(V^{\star},E^{\star})$ be the ground-truth service-level fault propagation
graph and $G=(V,E)$ the agent's predicted graph from its final result
(\Cref{sec:method_annotation}). We compute
\begin{equation}
  \mathrm{Node\,F1} = \frac{2|V \cap V^{\star}|}{|V| + |V^{\star}|},
\end{equation}
\begin{equation}
  \mathrm{Edge\,F1} = \frac{2|E \cap E^{\star}|}{|E| + |E^{\star}|}.
\end{equation}
Node\,F1 measures service-set recovery; Edge\,F1, directed propagation-link
reconstruction. In \Cref{tab:rq1_main}, rds, Node\,F1, and Edge\,F1 are split by $\mathrm{Acc}@1$ outcome ($\checkmark$: correct; $\times$: incorrect; avg: all cases).

\paragraph{Causal-chain depth.}
We stratify cases by the \textbf{causal-chain depth} $d^{\star}$, the
shortest-path length from the root cause to the
SLO-alarming service along $G^{\star}$.

\subsection{Behavior and Intent Analysis}\label{sec:method_behavior}

The diagnostic goal is to localize a root cause. The services an agent queries at each round reveal whether it is advancing toward that cause along the fault-impact surface ($V^{\star}$, $E^{\star}$), and each action's SQL exposes the diagnostic intent behind that advance. We accordingly build two process views
over the trajectories: a \emph{behavior} view, tracking how each round navigates
the fault-impact surface and uses surfaced evidence, and an \emph{intent} view,
typing each SQL action by its diagnostic intent.

\subsubsection{Impact-surface navigation and evidence utilization}\label{sec:method_transitions}

We define three per-round labels to characterize each trajectory: a \emph{primary service} that tracks where the investigation stands, a \emph{navigation transition} that captures its direction, and an \emph{evidence-utilization} label that records whether it acts on returned evidence.

\paragraph{Per-round primary service.}
Each round touches several services, so to track this advance at round granularity
we reduce it to a single \emph{primary service}: foremost the round's service
nearest a root cause, capturing how far the round advanced. Among equally near
services, it is the newly visited one, which marks a new investigation direction,
or else the most frequently visited, where the round's investigation has
concentrated. Concretely, we draw candidates from the round's own services (its
\emph{action services}, or its result services when the round issues no query) and
rank them lexicographically by (i) distance from the nearest root cause along
$G^{\star}$, with off-path services taking a large penalty; (ii) unvisited before
visited; then (iii) higher cumulative visit count (both relative to earlier
rounds). Key (i) sets the round's status (\emph{on-path} if any candidate is
on-path, i.e., the round's own queries reach the propagation path, and
\emph{off-path} otherwise), and keys (ii) and (iii) break the remaining ties in
either case.

\begin{figure*}[t]
\centering
\includegraphics[width=\textwidth]{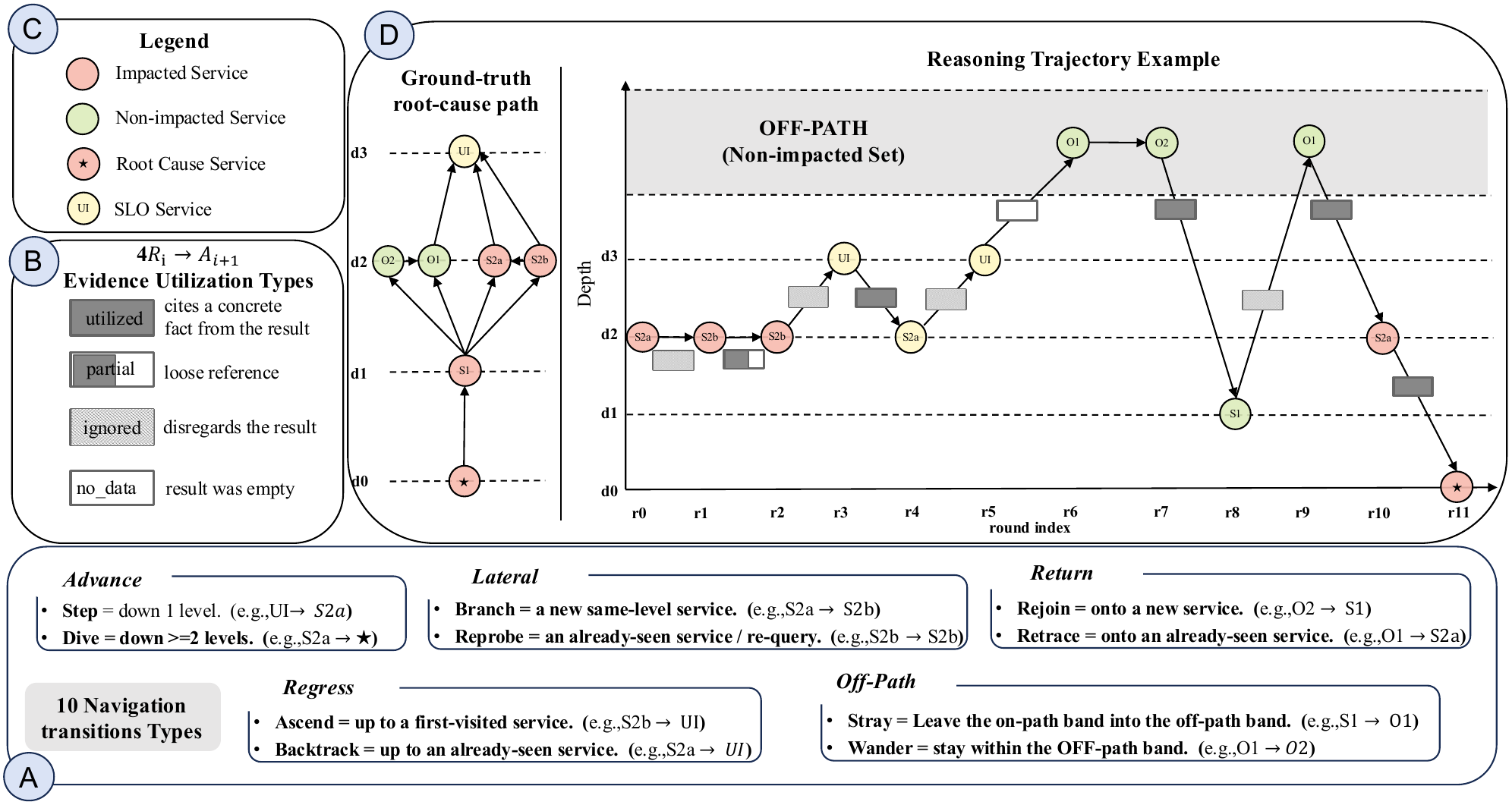}
\caption{\textit{Trajectory label vocabularies (RQ2 behavior view).} Each
reasoning round is reduced to a \emph{primary service} at some depth on the graph
(\Cref{sec:method_transitions}). \textit{Panel \textbf{A}}: the 10 navigation-transition
types, grouped into five families by how the investigation moves relative to a
root cause. \textit{Panel \textbf{B}}: the 4 evidence-utilization types, i.e., whether a
round acts on the previous round's on-path evidence. \textit{Panel \textbf{C}}: node legend
for panel D. \textit{Panel \textbf{D}}: a worked example plotting each round's
primary-service depth ($y$) over round index ($x$); the grey band is the off-path
region and the shaded box under each round marks its evidence-utilization type.
These labels are defined in \Cref{sec:method_transitions} and quantified in
\Cref{fig:rq2_transition_grid}.}
\label{fig:transition_defs}
\end{figure*}

\paragraph{Navigation transitions.}
For each round pair $(i, i{+}1)$ we label a \emph{navigation transition} from the
on-path status and distance-to-root of their two primary services, capturing
whether the investigation advances toward, stays level with, or moves away from a
root cause (full label set in \Cref{fig:transition_defs}).

\paragraph{Evidence utilization.}
Independently, an \emph{evidence-utilization} label compares round $i{+}1$'s
action services $A_{i+1}$ against the on-path subset of round $i$'s result services
$R_i$, capturing whether the next round acts on the decisive evidence the previous
one surfaced (\Cref{fig:transition_defs}).

\subsubsection{Intent taxonomy and sequence patterns}\label{sec:method_intent}

\begin{figure}[t]
\centering
\includegraphics[width=0.92\linewidth]{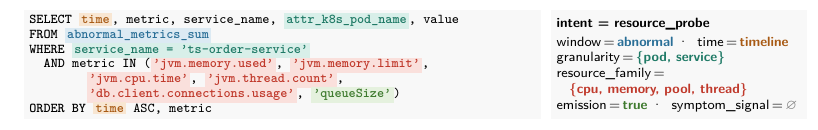}
\caption{\textit{Worked example of the intent/tag labeling applied to every agent SQL action in the study.} This abnormal-window metric query receives a single SRE \emph{intent} (\texttt{resource\_probe}) plus six deterministic \emph{tags}; color-matched highlights link each SQL fragment to the tag axis it sets (vocabulary in \Cref{tab:intent_taxonomy}).}
\label{fig:intent_tag_example}
\end{figure}

For the intent view, every SQL action receives one intent label and a
multi-valued set over six deterministic tag axes
(\Cref{tab:intent_taxonomy}; \Cref{fig:intent_tag_example}). A fixed LLM
classifier (\texttt{claude-opus-4.8}), following the LLM-as-judge
paradigm~\cite{zhuge2025agent,gu2024survey,li2024llmsasjudges}, labels
92{,}501 SQL actions by reading the compacted trajectory while anchoring on
the SQL operation and returned evidence rather than the agent's narrative.
The six tag axes are derived deterministically from the SQL text;
agreement and classifier accuracy are reported in \Cref{sec:method_reliability}.

\begin{table*}[t]
\centering
\caption{SRE diagnostic-intent and tag vocabulary used to label each agent action. The eleven intents (left) classify the diagnostic purpose of each SQL query; the six tag axes (right) are derived deterministically from the SQL text and describe its operational scope.}
\label{tab:intent_taxonomy}
\scriptsize
\setlength{\tabcolsep}{2pt}
\renewcommand{\arraystretch}{1.02}
\begin{tabularx}{\textwidth}{@{}p{0.04\textwidth}p{0.07\textwidth}p{0.14\textwidth}X@{}}
\toprule
\textbf{Axis} & \textbf{Family} & \textbf{Name} & \textbf{Definition} \\
\midrule
\multirow{11}{*}{Intent}
& \multirow{3}{*}{Symptom} & \texttt{symptom\_scan} & Compares service-level symptoms, including errors, latency, status codes, payload shape, and request volume. \\
& & \texttt{log\_inspect} & Inspects log records for errors, exceptions, stack traces, abnormal volume, and other textual symptoms. \\
& & \texttt{text\_search} & Searches telemetry text for service names, error patterns, fault keywords, and suspicious literals. \\
\cmidrule(lr){2-4}
& \multirow{3}{*}{Topology} & \texttt{trace\_follow} & Follows trace IDs, spans, and parent-child links to localize propagation along service calls. \\
& & \texttt{callgraph} & Queries caller-callee relations, span kinds, and call ratios to recover dependency direction and structure. \\
& & \texttt{network\_flow} & Inspects request flow, failures, connectivity, packet/drop signals, and service-to-service traffic. \\
\cmidrule(lr){2-4}
& \multirow{2}{*}{Resource} & \texttt{resource\_probe} & Tests resource hypotheses with metrics for named services, pods, nodes, JVMs, DBs, or runtimes. \\
& & \texttt{metric\_dump} & Retrieves broad metric values or snapshots to scan resource state before narrowing a hypothesis. \\
\cmidrule(lr){2-4}
& Liveness & \texttt{lifecycle\_probe} & Checks lifecycle/emission signals such as restarts, silence, span export, pod state, and JVM state. \\
\cmidrule(lr){2-4}
& \multirow{2}{*}{Inventory} & \texttt{metric\_enum} & Enumerates metric names, schemas, coverage, resource attributes, and measurement inventory. \\
& & \texttt{service\_enum} & Enumerates services, pods, nodes, tables, topology entities, and telemetry inventory. \\
\midrule
\multirow{6}{*}{Tag}
& \multirow{6}{*}{Derived} & \texttt{granularity} & Target level over which evidence is grouped, e.g., global, node, pod, service, endpoint, or trace. \\
& & \texttt{window} & Telemetry window compared by the query, e.g., abnormal, normal, or normal-abnormal contrast. \\
& & \texttt{time} & Temporal shape requested by the query, e.g., snapshot, timeline, or onset analysis. \\
& & \texttt{emission} & Whether the query references telemetry emission, silence, exporter, queue, or liveness fields. \\
& & \texttt{resource\_family} & Resource axis referenced by the query, e.g., CPU, memory, GC, pool, thread, or disk. \\
& & \texttt{symptom\_signal} & Symptom signal referenced by the query, e.g., error, status code, latency, payload, or volume. \\
\bottomrule
\end{tabularx}
\end{table*}

We summarize the intent view in two ways: the \emph{intent distribution}, giving
each experiment's frequency over the intents, and \emph{intent $n$-grams},
capturing contiguous subsequences that surface recurring motifs and
repeated-intent loops.

\subsection{Failure-Mode Coding and Mining}\label{sec:method_repair}

We code each failed trajectory against the expected evidence for its fault type,
yielding three failure families that differ in how the agent relates to decisive
evidence (\Cref{tab:rq3_failure_families}).

\subsubsection{Coding failures against expected evidence}\label{sec:method_failure}

A correct diagnosis must recover sufficient evidence for at least one anomalous
path from the observed symptom to a root cause on the fault-impact surface. Using
the RQ2 intent and tag vocabulary, we specify each fault's \emph{expected
evidence}, the typed queries that would confirm the relevant surface services and
dependencies, and read a failed trajectory as a shortfall: evidence the
agent omitted or misread, or reasoning untied to evidence.

For each failed trajectory we compare expected evidence against the agent's
actual queries and reasoning, drawing on its compacted trajectory, the
ground-truth causal context, and raw telemetry when needed.

We code deviations into three families (\Cref{tab:rq3_failure_families}):
\emph{protocol omission} (\texttt{OMIT}), evidence reachable in telemetry but never
queried; \emph{semantic misread} (\texttt{MIS}), retrieved evidence misinterpreted or
scoped to the wrong dependency; and \emph{general reasoning} (\texttt{GEN}), a
failure not tied to a single evidence unit.

Two authors jointly develop the codebook through open coding of failed
trajectories, letting the recurring failure codes emerge and refining them
until the codebook stabilizes; an LLM (\texttt{claude-opus-4.8}) assists
evidence retrieval~\cite{zhang2025agent,zhuge2025agent}, but every code is
verified by hand against the cited SQL, returned evidence, and the TrainTicket
service-call graph~\cite{trainticket}. Using the fixed codebook, the
two authors independently code all 154 failed \textsc{ThinkDepth.ai} trajectories (50 Sonnet, 104
Qwen); agreement is reported in \Cref{sec:method_reliability}.

\subsection{Annotation Reliability}\label{sec:method_reliability}

We assess the reliability of propagation-graph annotations,
SRE intent labels, and failure-mode annotations in RQ3 and RQ4.
For propagation graphs, intent labels, and RQ4 annotations,
we follow a shared review protocol: one annotator produces the
initial annotations with algorithmic or LLM assistance, as
appropriate; two reviewers independently accept or reject each
annotation; and a senior author adjudicates disagreements.
The agreement rates and Cohen's $\kappa$ reported for these
three tasks measure agreement between the reviewers' binary
acceptance decisions.

For all \faultGrandTotal{} propagation graphs, the reviewers
agree on 97.4\% of decisions ($\kappa{=}0.59$), with individual
acceptance rates of 96.8\% and 96.6\%.
For a stratified sample of 1{,}000 intent labels, they agree on
98\% ($\kappa{=}0.73$); after adjudication, 96\% of the original
labels are accepted.
For a stratified sample of 180 RQ4 failure-mode annotations
(15 per framework--backbone configuration), they agree on
97.2\% ($\kappa{=}0.53$); after adjudication, 95.6\% of the
original annotations are accepted.

For RQ3, after the codebook developed through open coding
(\Cref{sec:method_failure}) stabilizes, two authors independently
recode all 154 incorrectly localized trajectories using the
codebook. They assign identical code sets to
144 trajectories (93.5\%), measuring exact agreement at the
trajectory level.

\section{Characterization Results}\label{sec:experiments}

This section characterizes RCA agents along three deepening levels: RQ1 establishes outcome and cost, RQ2 explains the outcome gaps through behavior and intent, and RQ3 distills wrong diagnoses into recurring failure modes, which \Cref{sec:treatment} later turns into an intervention.

\subsection{RQ1: Effectiveness and Cost}\label{sec:rq1}

\begin{table*}[t]
\centering
\caption{\textbf{RQ1}: outcome-level comparison of all ten framework--backbone configurations on RCABench.
Edge\,F1 trails Node\,F1 by a wide margin, and Acc@1 falls as the causal chain deepens, a decline the Sonnet row largely escapes. Metrics and the $\checkmark$/$\times$ outcome split defined in \Cref{sec:method_defs}; Acc@1 and depth columns
in \%, F1 in $[0,1]$. Costs use the model prices in \Cref{sec:prelim_models}.}
\label{tab:rq1_main}
\footnotesize
\setlength{\tabcolsep}{0.9pt}
\resizebox{\textwidth}{!}{%
\begin{tabular}{@{}llcccccccccccccccc@{}}
\toprule
& & & \multicolumn{3}{c}{Node\,F1}
& \multicolumn{3}{c}{Edge\,F1}
& \multicolumn{3}{c}{Rounds (\textbf{rds})}
& \multicolumn{2}{c}{Budget / case}
& \multicolumn{4}{c}{Acc@1 by depth $d^{\star}$}\\
\cmidrule(lr){4-6}\cmidrule(lr){7-9}\cmidrule(lr){10-12}\cmidrule(lr){13-14}\cmidrule(lr){15-18}
\textbf{Backbone model} & \textbf{Framework}
  & Acc@1
  & $\checkmark$ & $\times$ & avg
  & $\checkmark$ & $\times$ & avg
  & $\checkmark$ & $\times$ & avg
  & \textit{ktok} & \$
  & $d^{\star}{=}2$ & $d^{\star}{=}3$ & $d^{\star}{=}4$ & $d^{\star}{=}5$\\
\midrule
\multirow{9}{*}{Qwen}
& \textsc{mABC}       & 42.6 & 0.49 & 0.35 & 0.41 & 0.15 & 0.05 & 0.09 & -- & -- & -- & 163  & 0.04 & 53.3 & 36.4 & 40.4 & 42.9\\
 & \textsc{OpsAgent} & 45.8 & 0.81 & 0.52 & 0.65 & 0.61 & 0.26 & 0.42 & -- & -- & -- & 83 & 0.01 & 42.1 & 49.6 & 44.7 & 42.9\\
& \textsc{OpenRCA}    & 46.6 & 0.57 & 0.48 & 0.52 & 0.36 & 0.16 & 0.25 & 6.1  & 6.7  & 6.4  & 364  & 0.05 & 58.6 & 46.9 & 30.9 & 19.0\\
 & \textsc{GoS} & 48.2 & 0.77 & 0.51 & 0.64 & 0.57 & 0.25 & 0.40 & -- & -- & -- & 92 & 0.01 & 57.9 & 50.0 & 33.0 & 19.0\\
 & \textsc{RCLAgent V2} & 53.4 & 0.74 & 0.42 & 0.59 & 0.51 & 0.16 & 0.35 & 108.9 & 175.6 & 140.0 & 773 & 0.13 & 56.6 & 55.7 & 43.6 & 52.4\\
& \textsc{TaskWeaver} & 65.2 & 0.78 & 0.55 & 0.70 & 0.58 & 0.26 & 0.47 & 13.2 & 14.5 & 13.7 & 672  & 0.09 & 79.6 & 68.9 & 39.4 & 33.3\\
& \textsc{AIQ}        & 77.6 & 0.81 & 0.63 & 0.77 & 0.61 & 0.35 & 0.55 & 50.0 & 55.1 & 51.1 & 1516 & 0.19 & 86.8 & 82.0 & 59.6 & 42.9\\
& \textsc{ClaudeCode} & 79.6 & 0.81 & 0.58 & 0.76 & 0.65 & 0.28 & 0.57 & 51.3 & 62.7 & 53.6 & 1512 & 0.18 & 89.5 & 79.8 & 68.1 & 57.1\\
\cdashline{2-18}
& \textsc{ThinkDepth.ai} & 79.2 & 0.83 & 0.59 & 0.78 & 0.66 & 0.33 & 0.59 & 43.1 & 50.7 & 44.7 & 1744 & 0.21 & 85.5 & 82.0 & 66.0 & 57.1\\
\midrule
Sonnet & \textsc{ThinkDepth.ai} & 90.0 & 0.82 & 0.56 & 0.79 & 0.67 & 0.37 & 0.64 & 30.5 & 38.2 & 31.2 & 1465 & 4.64 & 88.2 & 88.6 & 96.8 & 85.7\\
\bottomrule
\end{tabular}%
}
\end{table*}

\paragraph{Framework and Model Comparison}
\Cref{tab:rq1_main} compares agents along two axes. Across frameworks, effectiveness is shaped by investigative architecture: the
agents that run an open-ended, adaptive investigation, requerying telemetry as
evidence emerges (\textsc{ThinkDepth.ai}
\acAtOneThinkdepthaiQwen{}, \textsc{AIQ} \acAtOneAiqQwen{},
\textsc{ClaudeCode} \acAtOneClaudecodeQwen{}), hold the top band in
a narrow range, while the more constrained designs trail:
\textsc{TaskWeaver} (\acAtOneTaskweaverQwen{}) follows a fixed plan; \textsc{RCLAgent~V2} ($53.4\%$) spins up one sub-agent per
trace span and checks each in isolation; \textsc{GoS} ($48.2\%$) fixes
its hypotheses up front and, capped at three retrieval steps, tests only
the most credible one; \textsc{OpenRCA} (\acAtOneOpenrcaQwen{}) commits
as soon as it can emit an answer; \textsc{OpsAgent} ($45.8\%$) issues no
telemetry queries of its own, instead reasoning over a pre-extracted
feature set; and \textsc{mABC} (\acAtOneMabcQwen{}) does not iterate over
evidence at all. Notably, none of the top agents was built for RCA, yet
every agent built for it trails them: open-ended investigative frameworks
show a clear advantage on end-to-end RCA over those limited by a fixed
control flow or constrained data access.

Across backbone models, the
Sonnet \textsc{ThinkDepth.ai} arm reaches \acAtOneThinkdepthaiSonnet{} Acc@1, a
$10.8$-pp gain over Qwen. The top-band frameworks span only $2.0$\,pp
under the same model (\acAtOneAiqQwen{}--\acAtOneClaudecodeQwen{}), while the model
swap yields five times that gap, suggesting the backbone model as the dominant
lever.

\begin{findingbox}{Finding 1}
For end-to-end diagnostic tasks, adaptive, minimally constrained frameworks outperform more tightly orchestrated designs; among the top-band frameworks, the backbone model is the dominant lever.
\end{findingbox}

\paragraph{Causal-Chain Depth and Reconstruction Granularity}
Causal-chain depth serves as a proxy for diagnostic difficulty:
a longer chain demands more investigative steps and more
evidence to reach the root cause. Empirically, accuracy declines
accordingly. Nearly all Qwen frameworks lose ground from
depth~2 to depth~5 (\textsc{ClaudeCode} $89.5\to57.1\%$, \textsc{AIQ}
$86.8\to42.9\%$); on the matched \textsc{ThinkDepth.ai} pair, Qwen falls likewise
($85.5\to57.1\%$) while Sonnet holds flat ($85.7$--$96.8\%$), widening the
backbone-model gap on the hardest cases.

Acc@1 and surface reconstruction are positively coupled across
\Cref{tab:rq1_main}, yet the backbone-model swap lifts localization without
lifting the reconstruction ceiling. Weaker agents do not merely select
the wrong final service: they reconstruct less of the fault-impact
surface even on their correctly localized cases ($\checkmark$ Node\,F1
rises from $0.49$ for \textsc{mABC} to $0.83$ for \textsc{ThinkDepth.ai},
$\checkmark$ Edge\,F1 from $0.15$ to $0.66$). Even correct localizations,
however, rest on partial causal maps: $\checkmark$ Edge\,F1 never exceeds
$0.67$, and the model swap that raises Acc@1 by $10.8$\,pp leaves this
ceiling nearly untouched (Node\,F1 $0.82$ vs.\ $0.83$, Edge\,F1 $0.67$ vs.\
$0.66$). In every row, Edge\,F1 falls well below Node\,F1 (\textsc{mABC}:
$0.41$ vs.\ $0.09$; \textsc{ClaudeCode}: $0.76$ vs.\ $0.57$), and failure
widens this imbalance: from correct to incorrect runs, Node\,F1 drops by
$16$--$43\%$ while Edge\,F1 drops by $43$--$69\%$, so failed runs still
detect most anomalous services yet no longer chain them into a
propagation path. Identifying anomalous nodes requires only detecting
aberrant telemetry; recovering edges requires inferring how the
fault propagates between services, making edge reconstruction the harder
and more discriminative capability.

\begin{findingbox}{Finding 2}
Accuracy degrades with causal-chain depth, a decline the stronger backbone model largely escapes. Reconstructing the fault-impact surface is harder than selecting the correct final answer and remains partial even when localization succeeds; within the reconstruction, recovering directed propagation edges proves more demanding than identifying anomalous nodes.
\end{findingbox}

\paragraph{Reasoning Budget and Diagnostic Outcome}
Budget scales with investigative openness: the top-band agents sustain $45$--$54$
effective rounds, over $1500$\,ktok, and roughly \$0.2 per case, whereas the constrained
agents stop far sooner (under $700$\,ktok, at most \$0.09).
\textsc{RCLAgent~V2} is the exception: it runs a separate sub-agent for
each trace span, so its round counts grow with trace size rather than
with evidence-driven, open-ended search. Within
\textsc{ThinkDepth.ai}, Sonnet stops in fewer rounds than Qwen, locking onto decisive
evidence quickly where Qwen enumerates more broadly, though
Sonnet’s higher per-token price still dominates per-case cost.

Yet one pattern holds across the sequentially iterating agents: incorrect cases run more rounds than correct ones, by up to $11.4$ rounds. The extra rounds buy no partial credit either: incorrect runs trail correct ones on both F1 metrics, so the added budget reflects unproductive search rather than evidence accumulation.

\begin{findingbox}{Finding 3}
Reasoning budget varies across frameworks and backbone models, but within a framework more rounds do not yield better diagnoses: failed runs consume more rounds than successful ones, as unproductive search inflates cost without improving outcomes.
\end{findingbox}

\subsection{RQ2: Agent Behavior}\label{sec:rq2}
Having established outcome and cost differences, we now examine the diagnostic
process behind them. Because RQ2--RQ3 analyze SQL-call trajectories, they
focus on the four general-purpose frameworks and exclude the five
RCA-specific frameworks, which retain their native data access
(\Cref{sec:method_defs}).
We center on \textsc{ThinkDepth.ai}'s Qwen and Sonnet arms, the only matched
pair; the Qwen arm is representative because all Qwen SQL-call frameworks show similar
behavior and intent patterns.

\begin{figure*}[t]
\centering
\begin{subfigure}{\textwidth}
  \centering
  \includegraphics[width=0.82\linewidth]{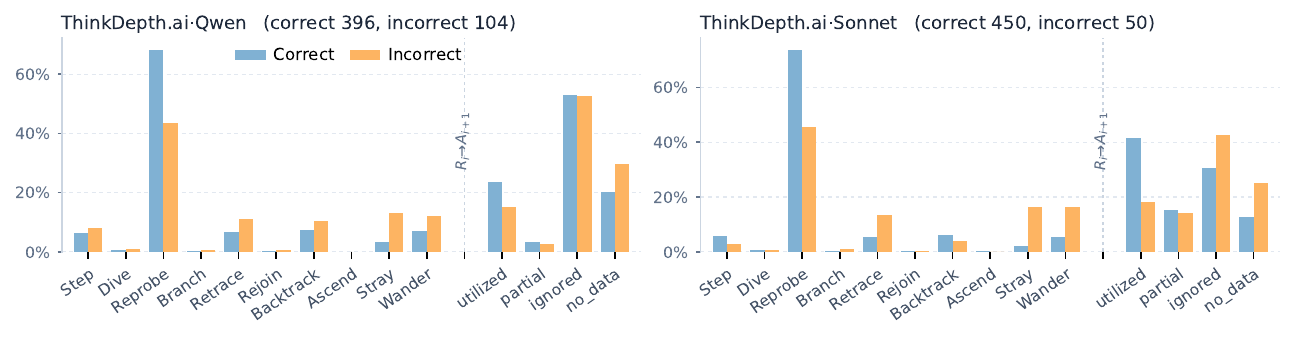}
  \caption{\textbf{Navigation and evidence-utilization label shares}, correct
  (blue) vs.\ incorrect (orange) runs per arm. In the navigation block (left),
  the correct--incorrect contrast is consistent across both arms; in the
  evidence-utilization block (right, after the dashed line), the two backbone
  models diverge sharply. Labels defined in \Cref{fig:transition_defs};
  bars macro-averaged per trajectory.}
  \label{fig:rq2_transition_grid}
\end{subfigure}

\vspace{3pt}

\begin{subfigure}{\textwidth}
  \centering
  \includegraphics[width=\linewidth,height=0.27\linewidth]{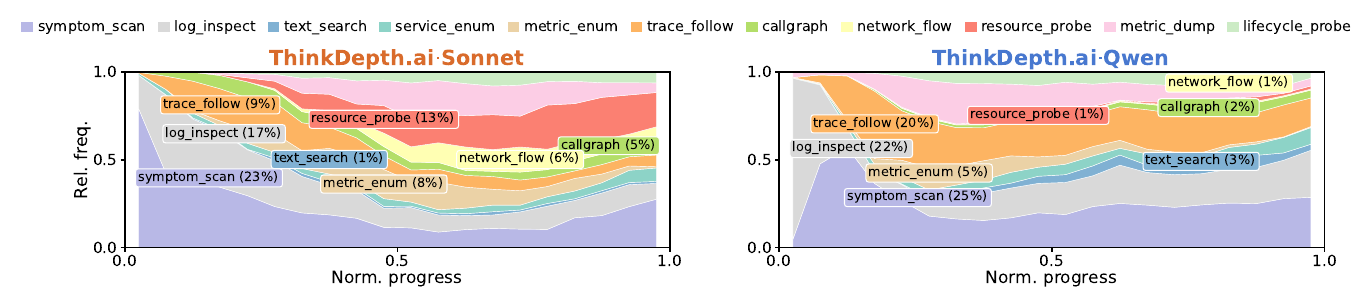}
  \caption{\textbf{SRE diagnostic-intent composition} over normalized trajectory
  progress. Both arms share a symptom-dominated opening phase; band shapes
  diverge from mid-trajectory onward, with the Sonnet arm shifting weight
  toward intent categories largely absent in Qwen. Band thickness is each
  intent's relative frequency (colors: the 11 intents of
  \Cref{tab:intent_taxonomy}); in-plot labels give overall shares.}
  \label{fig:rq2_move_stream}
\end{subfigure}
\caption{\textbf{RQ2}: behavior and intent analysis. What separates
correct from incorrect runs~(a) and what drives the backbone-model gap in
diagnostic strategy~(b).}
\label{fig:rq2_combined}
\end{figure*}

\paragraph{Surface Navigation and Evidence Grounding}\label{sec:rq2_transitions}
Correct runs ground each round in retrieved evidence; incorrect runs drift
off-path or ignore decisive clues (\Cref{fig:rq2_transition_grid}).
Specifically, correct runs lean on \texttt{Reprobe}, re-querying on-path
services at the same depth (higher than incorrect runs in both arms),
whereas incorrect runs oscillate on and off the surface (higher
\texttt{Stray}/\texttt{Wander} and \texttt{Retrace}). The two arms navigate in similar proportions.

Evidence grounding, not navigation volume, is where the two backbone models
diverge. Correct runs act on the evidence they retrieve (higher
\texttt{utilized}), whereas incorrect runs more often hit empty queries
(higher \texttt{no\_data}) and, in the Sonnet arm, ignore results (higher \texttt{ignored}). Sonnet acts on the prior result (\texttt{utilized} or \texttt{partial}) on $54.3\%$ of rounds versus Qwen's $24.9\%$. In effect, Sonnet's backtracks and advances are evidence-guided confirmations, while Qwen swings between shallow and deep nodes regardless of the returned evidence.

\begin{findingbox}{Finding 4}
Success and failure come down to process discipline: a successful diagnosis stays within the fault-impact surface and uses the evidence it retrieves, while a failing one wanders off or ignores decisive evidence.

\end{findingbox}

\paragraph{Intent Repertoire and Fault-Type Sensitivity}\label{sec:rq2_intent}
The two backbone models deploy qualitatively different diagnostic strategies, with
Sonnet pivoting mid-trajectory to resource- and topology-oriented intents where
Qwen persists in shallow trace and log loops
(\Cref{fig:rq2_move_stream}). Over the full eleven-intent profile, the
Sonnet arm is the least similar to every Qwen arm (cosine similarity
$0.907$--$0.928$), while the four Qwen frameworks cluster tightly
($0.947$--$0.989$).

Both arms open with a shared symptom-scanning
phase, after which their strategies diverge. Qwen stays trace- and log-bound: as
the trajectory advances it relies increasingly on fuzzy
\texttt{text\_search} over log and trace text and on \texttt{trace\_follow}
span enumeration, and although it inventories metrics via \texttt{metric\_enum}
and \texttt{metric\_dump}, it rarely drills into them. Sonnet runs the same
early symptom/log/trace survey but then pivots in mid-trajectory to exploring
the metric query space (\texttt{metric\_enum}), recovering dependency structure
with \texttt{callgraph}, and drilling named resources and flows via
\texttt{resource\_probe} (used in $82.8\%$ of its trajectories vs.\ $16.6\%$ for
Qwen) and \texttt{network\_flow}. This maps onto per-fault accuracy: Qwen trails
Sonnet most on categories requiring named-resource metrics (\textsc{Resource}
$75\%$ vs.\ $94\%$, \textsc{Code\,(JVM)} $66\%$ vs.\ $93\%$), while
\textsc{Network} nearly closes ($82\%$ vs.\ $89\%$) and \textsc{HTTP} is tied
($88\%$ vs.\ $87\%$).
Qwen further perseverates in shallow loops that Sonnet avoids: intent
5-grams such as sustained trace-following (\texttt{trace\_follow}$^{5}$, $0.96$ per
trajectory) and \texttt{log\_inspect}/\texttt{symptom\_scan} oscillation
recur far more in the Qwen arm than in Sonnet ($\leq\!0.12$ per trajectory), the sequence-level signature of its narrow repertoire.

\begin{findingbox}{Finding 5}
A richer diagnostic intent repertoire corresponds to broader, SRE-style search, while a narrow one keeps the agent in shallow loops over a limited slice of the evidence, producing fault-type-specific accuracy gaps that concentrate on the types requiring deeper evidence.
\end{findingbox}

\subsection{RQ3: Failure Taxonomy}\label{sec:rq3}

The behavioral patterns above show \emph{how} runs diverge; we now ask \emph{why}
wrong diagnoses occur.
Wrong diagnoses reduce to three evidence-handling failure families that differ in
how the agent relates to decisive evidence (\Cref{tab:rq3_failure_families}).

\begin{table*}[t]
\centering
\caption{\textbf{RQ3}: three evidence-handling failure families.
Each family addresses a different relationship between the agent and
decisive evidence: never gathered (\texttt{OMIT}), gathered but misread
(\texttt{MIS}), or overridden by ungrounded reasoning (\texttt{GEN}).}
\label{tab:rq3_failure_families}
\scriptsize
\setlength{\tabcolsep}{2.5pt}
\renewcommand{\arraystretch}{1.08}
\newcommand{\modeRowGray}{\cellcolor{gray!8}}
\begin{tabularx}{\textwidth}{@{}>{\raggedright\arraybackslash}p{0.070\textwidth}p{0.140\textwidth}>{\hsize=0.94\hsize\linewidth=\hsize}X>{\hsize=1.06\hsize\linewidth=\hsize}Xcc@{}}
\toprule
\textbf{Family} & \textbf{Mode} & \textbf{Definition} &
\textbf{Consequence} & \textbf{Sonnet} & \textbf{Qwen}\\
\midrule
\multirow{6}{=}[-14pt]{Protocol omissions}
& \modeRowGray \texttt{OMIT1}\newline Baseline gap
& \modeRowGray No comparison against a pre-incident healthy baseline.
& \modeRowGray Pre-existing signal misjudged as incident evidence.
& \modeRowGray $26.0\%$ & \modeRowGray $51.9\%$\\
& \texttt{OMIT2}\newline Edge untested
& No span-level test on the claimed caller--callee edge.
& Propagation edge asserted without span-level evidence.
& $54.0\%$ & $68.3\%$\\
& \modeRowGray \texttt{OMIT3}\newline Resource gap
& \modeRowGray No resource or lifecycle probe using concrete entity names.
& \modeRowGray Decisive resource or lifecycle signal remains unseen.
& \modeRowGray $34.0\%$ & \modeRowGray $63.5\%$\\
& \texttt{OMIT4}\newline Ranking gap
& No peer ranking to check if the signal is unique.
& Collateral spike shared across peers treated as uniquely causal.
& $2.0\%$ & $25.0\%$\\
& \modeRowGray \texttt{OMIT5}\newline Granularity gap
& \modeRowGray No drill-down to pod, endpoint, edge, or metric level.
& \modeRowGray Specific failing unit remains unidentified at fine granularity.
& \modeRowGray $18.0\%$ & \modeRowGray $16.3\%$\\
& \texttt{OMIT6}\newline Timing gap
& No check that proposed cause precedes the symptom.
& Causal claim lacks temporal corroboration.
& $4.0\%$ & $8.7\%$\\
\midrule
\multirow{3}{=}[-6pt]{Semantic misreads}
& \modeRowGray \texttt{MIS1}\newline Silence-as-health
& \modeRowGray Silence, flatness, or missing data read as healthy.
& \modeRowGray Suppressed or unreported fault incorrectly cleared as healthy.
& \modeRowGray $30.0\%$ & \modeRowGray $44.2\%$\\
& \texttt{MIS2}\newline Salience bias
& Loudest anomaly signal over-trusted without cause--symptom distinction.
& Noisy victim service selected as cause over quieter culprit.
& $42.0\%$ & $32.7\%$\\
& \modeRowGray \texttt{MIS3}\newline Semantics misread
& \modeRowGray Telemetry read with the wrong service-side semantics.
& \modeRowGray Fault direction reversed between caller and callee sides.
& \modeRowGray $86.0\%$ & \modeRowGray $61.5\%$\\
\midrule
\multirow{3}{=}[-6pt]{General reasoning}
& \texttt{GEN1}\newline Anchor lock
& Early hypothesis retained despite later conflicting evidence.
& Contradictory evidence discounted in the final answer.
& $28.0\%$ & $34.6\%$\\
& \modeRowGray \texttt{GEN2}\newline Schema-out claim
& \modeRowGray Causal state fabricated outside the prescribed set of states.
& \modeRowGray Unvalidatable out-of-schema state returned as root cause.
& \modeRowGray $82.0\%$ & \modeRowGray $20.2\%$\\
& \texttt{GEN3}\newline Friction stop
& Evidence channel dropped after tool error or empty return.
& Decisive evidence from that channel remains ungathered.
& $0\%$ & $12.5\%$\\
\bottomrule
\end{tabularx}
\vspace{2pt}
\begin{minipage}{0.96\textwidth}
\footnotesize
\emph{Note.} Percentages report within-model failure-pool prevalence
($n{=}50$ Sonnet, $n{=}104$ Qwen; all from \textsc{ThinkDepth.ai}); subcategories may overlap and do not sum to
100\%.
\end{minipage}
\end{table*}

\paragraph{Failure families and cross-model patterns}\label{sec:rq3_families}
The general-reasoning failures in \Cref{tab:rq3_failure_families} are untied to
any single signal: a model that cannot aggregate
scattered evidence or drill deep enough to discard distractors locks onto an early
anchor (\texttt{GEN1}); an overconfident one fills gaps by fabricating an
out-of-schema cause (\texttt{GEN2}); and one without the persistence to retry a
failed or empty query abandons the evidence channel altogether (\texttt{GEN3}).

The evidence-anchored families track the two backbone models' reasoning posture. Qwen fails predominantly by omission, Sonnet by misreading. Qwen has less working knowledge of the RCA
problem domain, so it stays redundantly confined to the most superficial, easily
found signals (\texttt{OMIT1}, \texttt{OMIT3}, \texttt{OMIT4}) and rarely redirects
its exploration from the evidence already gathered (\texttt{GEN1}). Sonnet has a
stronger ability to hypothesize new investigation directions from the evidence, but
its strong associative reach sometimes errs through hallucination (\texttt{GEN2}),
and its synthesis often falls short of pinning down the crux among contradictory
signals (\texttt{MIS2}, \texttt{MIS3}).

\begin{findingbox}{Finding 6}
Agent failures include three recurring families: decisive evidence is never gathered, is gathered but misread, or is overridden by ungrounded reasoning.
\end{findingbox}

Whether distilling these recurring modes into defenses can improve localization beyond the characterization setting remains to be tested.

\section{Intervention and Validation}\label{sec:treatment}

This section examines whether the RQ3 failure taxonomy can guide diagnostic interventions that generalize beyond the original characterization setting. We introduce \textsc{DiagGuard} (\Cref{sec:treatment_design}) and evaluate its localization gains across datasets, backbone models, and frameworks (\Cref{sec:validation}). We also examine whether the catalogued failure modes recur, how their prevalence changes under intervention, and whether framework-associated differences persist. Finally, component ablations assess the individual and combined contributions of \textsc{DiagGuard}'s two defenses (\Cref{sec:ablation}).

A held-out incident illustrates the intervention (\Cref{fig:case}). A disk-I/O fault on \texttt{tidb-tikv-0} manifests as downstream tail latency. The baseline \textsc{ThinkDepth.ai} agent anchors on a chronically noisy service and fails to investigate the storage tier, exhibiting several catalogued failure modes (\texttt{MIS2}, \texttt{OMIT1}, \texttt{OMIT3}, \texttt{OMIT5}, \texttt{OMIT6}, \texttt{GEN3}; \Cref{tab:rq3_failure_families}). With \textsc{DiagGuard}, the agent extends its investigation to the infrastructure tier and correctly identifies the faulty storage node.

\begin{figure}[tp]
\edef\pairedFigureHeight{\the\dimexpr0.238\linewidth\relax}
\begin{subfigure}[t]{0.49\linewidth}
\vspace{0pt}
\input{figures/treatment_overview}
\end{subfigure}\hfill
\begin{subfigure}[t]{0.49\linewidth}
\vspace{0pt}
\input{figures/case_study}
\end{subfigure}
\caption{(a) \textbf{The \textsc{DiagGuard} prototype.}
Two answer-agnostic defenses derived from the failure taxonomy
(\Cref{tab:rq3_failure_families}) augment the Diagnostician at three
temporal junctures. The \emph{Verifier} runs a diagnostic self-audit
pre-investigation and a soundness check pre-commitment; the \emph{Grounder}
supplies a search-space profile before the first query.
(b) \textbf{Motivating incident (Doubao / AIOps~2025).} An injected
disk-I/O fault on the TiKV storage node \texttt{tidb-tikv-0} surfaces only as
downstream tail latency; red-highlighted values mark abnormal-vs-normal shifts. The
baseline (red cross) anchors on the chronically noisy Ad service and never leaves
the service tier, whereas \textsc{DiagGuard} (green check) descends through the
infrastructure tier to the true root cause \texttt{tidb-tikv-0}
(\Cref{sec:treatment}).}
\end{figure}

\subsection{\textsc{DiagGuard} Design}\label{sec:treatment_design}

The taxonomy of \Cref{sec:rq3} doubles as a design specification:
its three failure families jointly inform the design of two
answer-agnostic defenses. \textbf{\textsc{DiagGuard}} wraps a
diagnostic framework's reasoning loop in these defenses at three
temporal junctures (\Cref{fig:treatment}), drawing on agentic
grounding~\cite{agentic-grounding} and inference-time
verification~\cite{test-time-verification} with rules traced to the
mined failure codes.

The core is the \emph{Diagnostician}, the unmodified
diagnostic framework, whose standalone
run serves as the baseline in \Cref{sec:validation}.
The \textbf{Verifier} injects a diagnostic self-audit before the
Diagnostician begins its investigation, preventing the \texttt{OMIT}
and \texttt{MIS} families (evidence routinely skipped or
misread) from going unchecked; when the Diagnostician drafts a verdict,
the Verifier injects a soundness check that curbs the \texttt{GEN}
family by requiring the agent to reconcile its answer with the evidence
already gathered. The \textbf{Grounder} interposes a
data-grounding layer before the Diagnostician writes its first query,
replacing passive schema inspection with a search-space profile that
surfaces overlooked diagnostic axes (\texttt{OMIT}) and anchors the
agent's queries in actual data contents (\texttt{MIS}). Both defenses
are integrated into each framework through prompts and tools, with rules
fixed across validation settings.

\subsection{RQ4: Intervention Effectiveness and Failure-Mode Recurrence}\label{sec:validation}
\begin{figure}[t]
\centering
\begin{subfigure}[t]{0.30\linewidth}
  \centering
  \includegraphics[width=\linewidth]{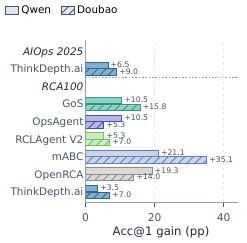}
  \caption{Acc@1 improvements over corresponding baselines}
  \label{fig:rq4_gain}
\end{subfigure}\hfill
\begin{subfigure}[t]{0.30\linewidth}
  \centering
  \includegraphics[width=\linewidth]{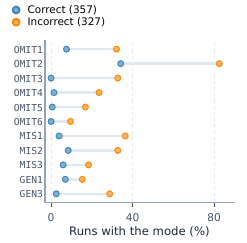}
  \caption{Failure-mode prevalence by baseline run outcome}
  \label{fig:rq4_modes_outcome}
\end{subfigure}\hfill
\begin{subfigure}[t]{0.36\linewidth}
  \centering
  \includegraphics[width=\linewidth]{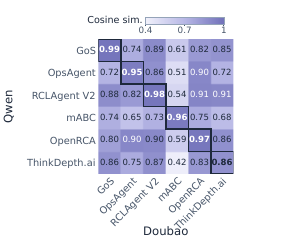}
  \caption{Failure-mode similarity across frameworks and backbones}
  \label{fig:rq4_modes_profile}
\end{subfigure}
\caption{\textbf{RQ4}: accuracy gains with \textsc{DiagGuard} and patterns of failure modes.
(a)~$\mathrm{Acc}@1$ improvement over baseline, in percentage points. Solid bars denote Qwen; hatched bars denote Doubao. All configurations improve with \textsc{DiagGuard}.
(b)~Percentage of correct and incorrect baseline runs exhibiting each failure mode, calculated separately for the two outcome groups. Every mode is more prevalent among incorrect runs.
(c)~Cross-backbone similarity of framework failure profiles. Each profile contains the prevalence of each failure mode among that framework's incorrect baseline runs. Rows represent frameworks using Qwen; columns represent frameworks using Doubao. Cell values indicate cosine similarity, with diagonal cells comparing the same framework across backbones. Failure-mode prevalence is more similar across backbones within the same framework than across frameworks.
Failure modes are defined in \Cref{tab:rq3_failure_families}.}
\label{fig:rq4_failure_modes}
\end{figure}

We evaluate whether \textsc{DiagGuard} improves localization beyond the characterization setting and whether the catalogued failure modes recur on RCA100. The evaluation covers \textsc{ThinkDepth.ai} on AIOps~2025 and six frameworks on RCA100, each under Qwen and Doubao (fourteen configurations; \Cref{fig:rq4_gain}). We apply the fixed RQ3 codebook to all 1,368 RCA100 runs (684 baseline and 684 \textsc{DiagGuard}) to examine mode recurrence, prevalence changes, and framework-associated failure profiles.

\paragraph{Generalization of intervention effectiveness}
With fixed intervention rules, \textsc{DiagGuard} improves $\mathrm{Acc}@1$ in all fourteen configurations. On AIOps~2025, \textsc{ThinkDepth.ai} gains $6.5$ and $9.0$ percentage points under Qwen and Doubao, respectively. Across the twelve RCA100 configurations, the median gain is $10.5$ percentage points (range: $3.5$--$35.1$).
The gains span graph-structured reasoning (\textsc{GoS}, $+10.5$/$+15.8$~pp under Qwen/Doubao), experience accumulation (\textsc{OpsAgent}, $+10.5$/$+5.3$), and general-purpose investigation (\textsc{ThinkDepth.ai}, $+3.5$/$+7.0$). Multi-agent orchestration also benefits: \textsc{mABC} shows the largest gains ($+21.1$/$+35.1$), from baseline accuracies of $35.1\%$ and $3.5\%$.
On RCA100, Node\,F1 and Edge\,F1 for the final reported propagation graphs each improve in eleven of twelve configurations. Their median increases across all twelve configurations are $0.06$ and $0.10$, respectively.

\begin{findingbox}{Finding 7}
With fixed intervention rules, \textsc{DiagGuard} improves localization accuracy in all fourteen configurations across datasets and backbone models, spanning diagnostic frameworks from constrained search or data-acquisition workflows to end-to-end, open-ended investigation.
\end{findingbox}

\paragraph{Failure-mode recurrence on RCA100}
All eleven evaluated modes (excluding \texttt{GEN2}, which framework constraints make incomparable) recur in baseline RCA100 runs, supporting their applicability beyond the original setting. The codes describe process defects that can occur even when localization is correct. Nevertheless, every mode is more prevalent in incorrect than correct baseline runs (\Cref{fig:rq4_modes_outcome}). Incorrect and correct baseline runs exhibit $3.29$ and $0.71$ distinct modes on average, respectively. Under \textsc{DiagGuard}, the corresponding means are $2.82$ and $0.63$, showing that the association with incorrect localization persists.

\paragraph{Changes in failure-mode prevalence}
Under \textsc{DiagGuard}, the mean number of distinct modes per run decreases from $1.94$ to $1.39$, with reductions in all twelve RCA100 configurations. Across 684 runs per condition, seven modes show statistically significant prevalence reductions. The largest decreases occur in \texttt{OMIT2} ($57.3\%$ to $44.7\%$) and \texttt{GEN3} ($15.1\%$ to $4.7\%$). \texttt{MIS3} (Semantics misread) does not decrease significantly under \textsc{DiagGuard} ($11.8\%$ to $13.6\%$). One possible explanation is that the intervention improves evidence acquisition without fully addressing errors in interpreting the retrieved evidence.

\paragraph{Framework-associated failure profiles after intervention}
We next examine whether framework-associated failure patterns persist after intervention. For each framework--backbone configuration, we compute an eleven-mode prevalence vector over its incorrect runs and compare vectors using cosine similarity. Each entry is the fraction of incorrect runs exhibiting the corresponding mode. Mean cross-backbone similarity is higher within the same framework than across frameworks under both baseline ($0.950$ vs.\ $0.767$; \Cref{fig:rq4_modes_profile}) and \textsc{DiagGuard} ($0.897$ vs.\ $0.720$). These comparisons indicate that framework-associated differences remain observable across the two evaluated backbones after intervention.

\begin{findingbox}{Finding 8}
All eleven evaluated modes recur on RCA100, and \textsc{DiagGuard} significantly reduces the prevalence of seven. Under both conditions, failure-mode profiles are more similar, on average, across backbones within the same framework than across frameworks.
\end{findingbox}

\subsection{RQ5: Component Contribution}\label{sec:ablation}
\Cref{tab:treat_ablation} ablates the two defenses on AIOps~2025
(Doubao, 400 incidents, five runs). Each helps on its own by a
comparable margin: the Grounder alone (\emph{w/o Verifier}) raises $\mathrm{Acc}@1$ by
$+4.5$ and the Verifier alone (\emph{w/o Grounder}) by $+4.1$, with the full \textsc{DiagGuard}
reaching $+9.0$ and the same ordering at $\mathrm{pass}@3$ and $\mathrm{pass}@5$. Both defenses improve accuracy individually, and combining them yields a larger gain than either alone, approximately matching the sum of their individual $\mathrm{Acc}@1$ gains.

\begin{table}[t]
\centering
\caption{\textbf{RQ5}: component ablation of \textsc{DiagGuard} on AIOps~2025 (Doubao, 400 incidents). The $\Delta$ column shows near-additive gains that do not track the budget columns (rds, ktok, \$): the costliest arm gains least, and the full stack gains most while spending less than it. $\mathrm{Acc}@1$ is mean$\pm$sd over five runs.}
\label{tab:treat_ablation}
\scriptsize
\setlength{\tabcolsep}{2.4pt}
\renewcommand{\arraystretch}{1.15}
\begin{tabular}{@{}lrrrrrrr@{}}
\toprule
& \multicolumn{2}{c}{$\mathrm{Acc}@1$} & $\mathrm{pass}@3$ & $\mathrm{pass}@5$
 & \multicolumn{3}{c}{Budget / incident}\\
\cmidrule(lr){2-3}\cmidrule(lr){4-4}\cmidrule(lr){5-5}\cmidrule(lr){6-8}
\textbf{Configuration} & \% & $\Delta$ & \% & \%
 & \textit{rds} & \textit{ktok} & \textit{\$}\\
\midrule
Baseline (w/o both)      & 43.5$_{\pm0.6}$ & ---   & 56.9 & 62.3 & 14.1 & 176.8 & 0.09\\
\quad{}w/o Grounder      & 47.6$_{\pm1.6}$ & $+4.1$ & 61.3 & 65.8 & 19.2 & 343.8 & 0.17\\
\quad{}w/o Verifier      & 48.0$_{\pm0.9}$ & $+4.5$ & 62.0 & 67.3 & 16.3 & 212.1 & 0.10\\
\textbf{\textsc{DiagGuard} (full)} & \textbf{52.5}$_{\pm1.1}$ & $\mathbf{+9.0}$ & \textbf{67.1}
 & \textbf{73.0} & 18.7 & 318.5 & 0.16\\
\bottomrule
\end{tabular}
\end{table}

\section{Threats to Validity}\label{sec:threats}

\paragraph{External validity.}
Findings from a single microservice system may not generalize. We therefore
evaluate on two additional systems with different topologies (AIOps~2025 and
RCA100) and compare across backbone models and frameworks; the failure modes
catalogued on RCABench recur on RCA100, suggesting that the observed patterns
are not system-specific. All three systems are open source and may appear in
pretraining data, but each framework and \textsc{DiagGuard} comparison holds the
backbone fixed, so such leakage affects compared arms equally.

\paragraph{Internal validity.}
We include all publicly available state-of-the-art RCA frameworks we could
execute, but some components are unreleased or incompatible with API-served
backbones (e.g., \textsc{OpsAgent}'s PPO self-evolution), which may understate
those frameworks. To ensure fair comparison, all frameworks run on the same
cases under the same backbone with the same metrics, while each RCA-specific
framework retains its native data access. Another threat is potential bugs in
our implementation; we address this through extensive testing and by releasing
all code, data, and scripts for independent verification.

\paragraph{Construct validity.}
Several labels require human judgment, which may introduce bias. Two annotators
independently label each task, with a senior annotator resolving disagreements
(93.5--98\% agreement; \Cref{sec:method_reliability}); we release the
annotation records and codebook for reproduction. Additionally, standard
metrics such as Top@K and MRR may not fully capture the practical
interpretability needs of an operator; we acknowledge this limitation and scope
our evaluation to automated localization accuracy.

\paragraph{Conclusion validity.}
A single run may not capture run-to-run variation. We repeat the AIOps~2025
ablation five times, reporting mean$\pm$sd and pass@$k$
(\Cref{tab:treat_ablation}), and draw conclusions from the consistent gain
across all fourteen configurations rather than from any single comparison.

\section{Conclusion}\label{sec:conclusion}

We presented a process-level study of how LLM agents reach a root-cause
diagnosis, not merely whether it is correct. Using manually validated fault
propagation paths, we analyzed 5{,}000 trajectories from ten
framework--backbone configurations. Correct localization often masks
incomplete reconstruction of fault propagation, particularly the links between
affected services. Detailed analysis of four frameworks with comparable SQL
actions shows that successful investigations tend to stay on the fault-impact
surface and use the evidence they retrieve, while failed investigations more
often drift off the surface or leave decisive evidence unused. A broader range
of diagnostic intents is associated with deeper investigation. We identify
three recurring failure families: protocol omissions, semantic misreads, and
general-reasoning failures.

\textsc{DiagGuard}, a prototype developed from these failure modes, improves
localization accuracy in all fourteen tested configurations across two
additional benchmarks, two backbone models, and six frameworks, with gains of
3.5 to 35.1 percentage points. The evaluated failure modes recur across
settings, and \textsc{DiagGuard} reduces their mean count per run. Even after
intervention, the distributions of failure modes remain, on average, more
similar across backbone models within the same framework than across
frameworks. Together, these findings support assessing fault propagation paths
and evidence use alongside localization accuracy, and using recurring failures
to guide improvements to individual frameworks. Future work will use annotated
paths and failure labels to guide agent training and extend the evaluation to
production incidents.

\bibliographystyle{ACM-Reference-Format}
\bibliography{ref}

\end{document}